\documentclass[12pt]{iopart}

\usepackage{hyperref}
\usepackage{graphicx}
\usepackage{rotating}
\usepackage{array}

\def\be{\begin{equation}}
\def\ee{\end{equation}}
\def\bea{\begin{eqnarray}}
\def\eea{\end{eqnarray}}

\def\ie{{\it i.e.\,}}
\def\etal{{\it et al.   }}

\def\hcal{\mbox{$\cal H\,$}}

\def\<{\langle}
\def\>{\rangle}

\begin{document}

\title{Phase Space Evolution and Discontinuous Schr\"odinger Waves}
\author{E. Sadurn\'i}
\address{Institut f\"ur Quantenphysik, Universit\"at Ulm, Albert-Einstein Allee 11 89081 Ulm - Germany.}

\ead{esadurni@uni-ulm.de}

\begin{abstract}

The problem of Schr\"odinger propagation of a discontinuous wavefunction $-$ diffraction in time $-$ is studied under a new light. It is shown that the evolution map in phase space induces a set of affine transformations on discontinuous wavepackets, generating expansions similar to those of wavelet analysis. Such transformations are identified as the cause for the infinitesimal details in diffraction patterns. A simple case of an evolution map, such as $SL(2)$ in a two-dimensional phase space, is shown to produce an infinite set of space-time trajectories of constant probability. The trajectories emerge from a breaking point of the initial wave.

\end{abstract}

\maketitle

\section{Introduction}

What are the consequences of breaking a Schr\"odinger wave? In a seminal paper \cite{mosh}, Moshinsky studied the evolution problem of a quantum-mechanical wave blocked by a shutter and identified his results with Fresnel diffraction. Since then, the so-called 'diffraction in time' has been widely studied \cite{zeilinger}, and experimentally pursued \cite{nussensveig}. Needless to say, the spatial side of the analogy (in the paraxial approximation \cite{hecht}) has a long tradition of its own, mostly in the context of electromagnetic theory \cite{hannay, nye}. 

It is rarely recognized, however, that the discontinuities of the initial condition are the true origin of Schr\"odinger diffraction as well as the peculiar intensity patterns accompanying the effect. The present paper delves into the subject by revealing its mathematical structure through the use of symmetries and self-similar relations obeyed by wave functions. It is shown that the afore mentioned patterns in space and time can be explained in a very general framework dealing with the evolution of discontinuous initial conditions. The role of symmetry in these quantum-mechanical problems will be crucial, as its influence in the evolution of the system shall give rise to non-classical trajectories of constant probability.

Although diffractive effects are described by old and simple wave theories, the underlying explanation of the complexity in the emerging patterns is a subject of current discussion \cite{muga} and the resulting intricacies are the center of attention of other studies in connection with fractality \cite{berry}. It is worth to mention that the study of discontinuities in quantum-mechanical wave functions is also motivated by current physical applications ranging from molecular Talbot interferometers \cite{bill} to matter-wave optics based on cold atoms \cite{phillips, turlapov}.

Structure of the paper: In the following section, we study the old problem of diffraction by edges, in order to show all the relevant features of patterns by means of elementary techniques, including self-similarity and caustics. In section 3 we present a generalization of the method and illustrate it with a few examples in section 4. These include the diffraction of a square packet in a parametric harmonic oscillator, the evolution of discontinuities under non-linear canonical evolution and their relation to wavelet expansions. As a final example, the evolution of a discontinuous packet under the Gross-Pitaevskii equation is shown numerically. In section 5 we give a brief conclusion. 

\section{Diffraction by edges}

\begin{figure} \begin{center} \includegraphics[scale=0.5, angle=270]{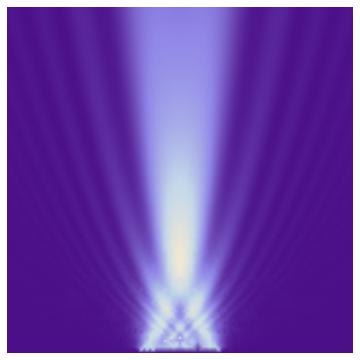}  \end{center} \caption{\label{cero} Probability density as a function of space and time. Two regions can be distiguished: 1) the near zone or short-time regime, where infinite oscillations and a main lobe can be identified, and 2) the far zone or long-time regime, where the wave function spreads} \end{figure}

It is usually considered that, in the paraxial approximation, the problem of diffraction by edges in the $x,z$ plane can be described by a plane wave in $z>0$ propagating parallel to the $z$ axis, a discontinuous opaque screen placed at $z=0$ parallel to the $x$ axis and the corresponding solution at $z>0$, subject to the boundary conditions mentioned before. The solution at $z>0$ can be found by propagating the wave at $z=0$, which is taken as the original plane wave but forced to vanish at the blocking screen \ie a discontinuous initial condition.  
In the following we solve the corresponding Schr\"odinger propagation problem in natural units ($\hbar=1=m$) using the time variable $t$ instead of $z$. We have
\bea
\left[-\frac{1}{2}\nabla^2 + V \right] \psi(x,t)= i \frac{\partial \psi(x,t)}{\partial t}, \qquad \psi(x,0) = \psi_0(x)
\label{0.1}
\eea
where $\psi_0$ has, in principle, an arbitrary number of discontinuities. See the single-slit example in figure \ref{cero}.

Let us consider a variant of the problem treated by Moshinsky \cite{mosh}, revisited by Nussenzveig \cite{nussensveig}. Consider the intial condition

\bea
\psi(x,t=0) = \theta(1/2-|x|)
\label{uno}
\eea
corresponding to a square packet of unit width. The wave function at time $t$ can be obtained via the free propagator \cite{grosche}, which is a Gaussian of the form

\bea
K(x,x';t,0) = \sqrt{\frac{1}{2i \pi t}} \exp \left( \frac{i (x-x')^2}{2t} \right).
\label{prop}
\eea
The wavefunction reads

\bea
\fl \psi(x,t)
 = \frac{1}{2} \mbox{Erf}\left[-e^{-i\pi/4} (x-1/2)/\sqrt{2t}\right] - \frac{1}{2} \mbox{Erf}\left[-e^{-i\pi/4} (x+1/2)/\sqrt{2t}\right] .
 \label{tres}
 \eea
In this form we can obtain the intensity pattern in space and time, as shown in figure \ref{cero}. It is worth to mention that the properties of such a pattern can be inferred from the Cornu spiral, as it was pointed out long ago \cite{mosh}. However, the applicability of this geometrical picture is restricted to the appearence of Fresnel integrals in the solution. More general initial conditions demand the use of more powerful methods.

\subsection{Self-similarity in single-slit diffraction using special functions}

\begin{figure}[h!]
\begin{center}
\begin{tabular}{ccc}
\includegraphics[scale=0.5]{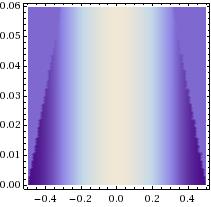} &
\includegraphics[scale=0.5]{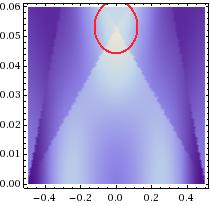} &
\includegraphics[scale=0.5]{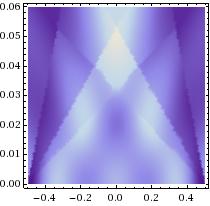} \\
\includegraphics[scale=0.5]{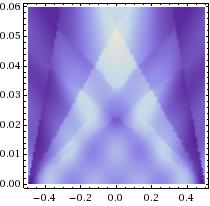} &
\includegraphics[scale=0.5]{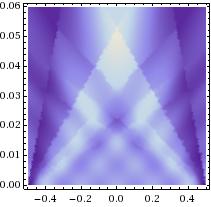} &
\includegraphics[scale=0.5]{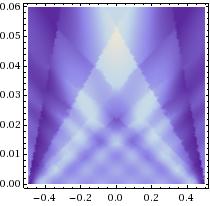} 
\end{tabular}
\end{center}
\caption{ \label{fig:once} Replication process using square packets as building blocks. The red circle indicates the pattern to be replicated.}
\end{figure}

\begin{figure}[h!]
\begin{center}
\begin{tabular}{ccc}
\includegraphics[scale=0.5]{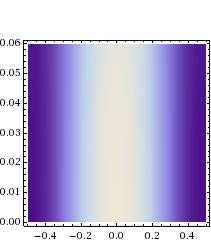} &
\includegraphics[scale=0.5]{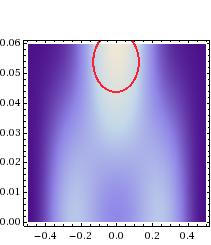} &
\includegraphics[scale=0.5]{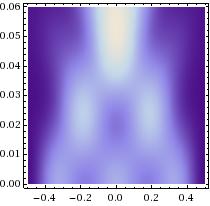} \\
\includegraphics[scale=0.5]{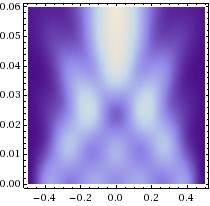} &
\includegraphics[scale=0.5]{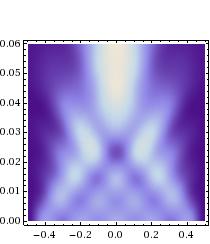} &
\includegraphics[scale=0.5]{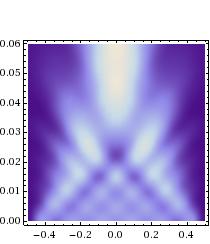} 
\end{tabular}
\end{center}
\caption{ \label{fig:doce} Replication process using gaussian packets as building blocks. The red circle indicates the pattern to be replicated.}

\end{figure}

\begin{figure}[h!]
\begin{center}
\begin{tabular}{ccc}
\includegraphics[scale=0.35]{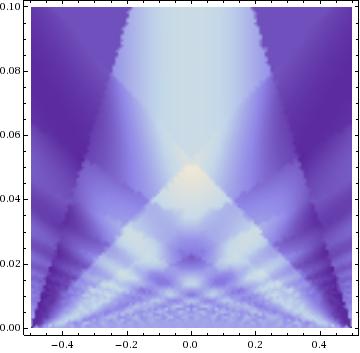} &
\includegraphics[scale=0.35]{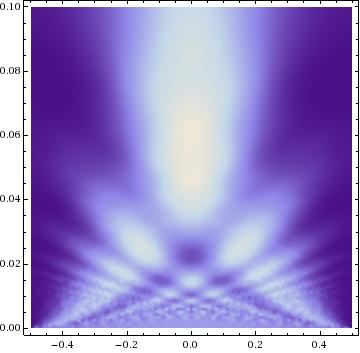} &
\includegraphics[scale=0.35]{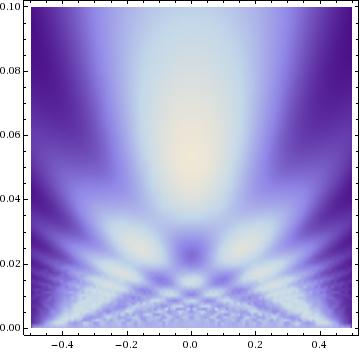} 
\end{tabular}
\end{center}
\caption{ \label{fig:trece} A comparison between the pattern obtained through the replication formula and the numerical calculation using Moshinsky functions. The expansions contain 30 terms.}

\end{figure}

\begin{figure}[h!]
\begin{center}
\begin{tabular}{ccc}
\includegraphics[scale=0.35, angle=90]{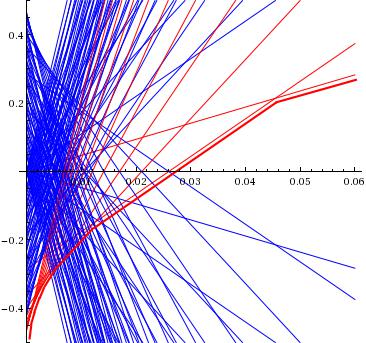} & \includegraphics[scale=0.35]{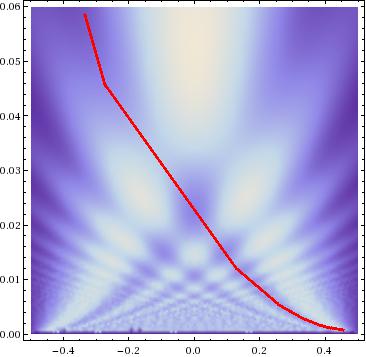} & \includegraphics[scale=0.35]{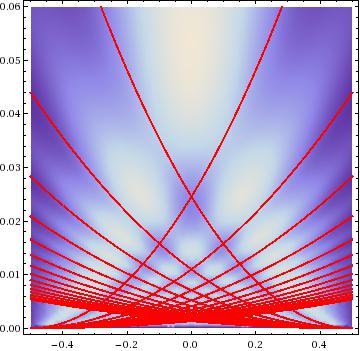}
\end{tabular}
\end{center}
\caption{The emergence of caustics as the culprits for the constant intensity lines in the pattern. As indicated in the text, the caustics can be computed explicitly in the form of parabolas}
\label{fig:diez}
\end{figure}

The origin of a self-similar pattern near the edges of the initial condition can be described alternatively by a {\it replication formula\ }, \ie a relation describing the wave function in terms of itself. Using a trigonometric expansion of the initial condition

\bea
\theta(1/2-|x|) = \frac{1}{2} + \frac{2}{\pi} \sum_{n=0}^{\infty} \frac{1}{2n+1} \sin \left( \pi (2n+1) (x+1/2) \right)
\label{4}
\eea
and applying the Gaussian propagator to this expansion, we can prove in a very simple way the identity

\bea
\mbox{Erf}\left[ e^{-i 3\pi/2}(1/2-x)/\sqrt{t}\right] = \sum_{n=0}^{\infty} C_n e^{i \xi_n(x,t)} \mbox{Erf}\left[ \eta_n(x,t)\right]
\label{aa}
\eea
where we have used the definitions
\bea
\quad k_n=\pi (2n+1), \quad C_n = 1/i k_n \\
\xi_n(x,t) = -k_n^2 t/2 + k_n(1/2 + x) \\
\eta_n(x,t) = e^{-i3\pi/2} \left( 1/2 - x - k_n t\right)/ \sqrt{t}. 
\label{bb}
\eea
Here we can infer the shape of the pattern near the edges by using a short time approximation of the functions on the r.h.s. of (\ref{aa}). We simply take the individual packets to be the initial conditions whose argument is displaced according to $\eta_n(x,t)$, that is

\bea
 \psi(x,t) &=& \sum_{n=0}^{\infty} C_n \left[e^{i \xi_n(x,t)} \theta (1/2-|\eta_n(x,t)|) + e^{i \xi_n(-x,t)} \theta (1/2-|\eta_n(-x,t)|) \right] \nonumber \\ &+& \frac{1}{2}\theta(1/2-|x|) + o(t^2)
\label{13}
\eea
where $O(t^2)$ is a small correction for short times. It is interesting to note that in this expansion, the emergence of the pattern in space time can be seen as the superposition of many individual square packets moving along the trajectories $\eta_n(\pm x,t)$, with plane wave factors propagating along $\xi_n(\pm x,t)$. The alternate signs indicate the contribution from each discontinuity or edge, while the index $n$ parameterizes the velocity. This velocity also alternates sign according to the source of the rays, propagating to the left from the right edge and vice versa. The individual functions preserving their integrity on the r.h.s. of (\ref{13}) can be regarded as building blocks, while the coefficients $C_n$ weight the contribution of each block. For the resulting probability density plots, see figures \ref{fig:once}, \ref{fig:doce}, \ref{fig:trece}. Furthermore, the building blocks can be replaced by any other localized shape such as that of a Gaussian or a tringular packet, and the resulting pattern generated by them will have the same features. See, in particular, figure \ref{fig:trece}. We distiguish the following properties:

\begin{itemize}
\item When the limit of the sums is truncated (say $n=n_{max}$), the pattern has a finite number of oscillations near the discontinuities. 

\item The second term $n_{max}=1$ shows already a main peak: The original pattern to be replicated. Increasing $n$ implies finer detail in the pattern near the edges.

\item Rays are described by $\xi_n(\pm x,t)= const$. The envelopes of such system of rays are caustics (Legendre transform) and can be obtained by eliminating $n$ from the condition $\partial_n \xi_n(x,t)=0$. It turns out that such envelopes are parabolas centered at the edges. See figure \ref{fig:diez}. 
\item Classical methods from geometric optics produce a non-classical result. To this respect we should point out that the trajectories of (approximately) constant probability obtained from the caustics should not be interpreted as classical trajectories. In free evolution, the latter are simply rays, while the former are parabolas. 
\item The short time approximation used in (\ref{13}) can always be improved. For example, in the case of initial space dependent phases, a velocity field given by the gradient of the phase produces additional motion on the initial packets or building blocks. Using the continuity equation for the initial probability density, one can infer the specific form of the motion and obtain the correct approximation.
\end{itemize}

\newpage

\section{Generalizations}

Many of the results above can be produced in more general settings. Now that we have revealed the origin of the patterns, we may consider intial conditions with an arbitrary number of discontinuities, as they can be written again in terms of individual square pulses

\bea
\psi_0(x) = \left[ \sum_{j} \theta(l_j/2-|x-x_j|) \right] \psi_0(x).
\label{1}
\eea
It is also possible to treat similar problems in many dimensions. Take, for example, the free evolution of a packet in three dimensions written as the product of pulses in each coordinate $x,y,z$. Even more general initial conditions with compact support can be considered, as long as we use the appropriate trigonometric expansions.

In this section we discuss not only the generalizations of the method, but also the role of symmetry in the evolution of discontinuities. Most of our treatment rests on the fact that a wave function can be written in terms of itself and on the possibility of representing its evolution as the superposition of wave packets which move and leave their mark in space-time, but preserve their integrity in the process.

\subsection{Replication formula from canonical evolution}

A discontinuous wave can be written in terms of itself using step functions. We resort to a trivial but remarkable identity for an individual pulse:

\bea
 \psi_0(x)=\theta(1/2-|x|) \psi_0(x). 
\eea
The evolution can be computed straightforwardly
\bea
\psi(x,t) &=& U_t \psi_0(x) = U_t \theta(1/2-|x|) U_{-t} U_t \psi_0(x) \\
&=& \theta(1/2-|\hat x (t)|) \psi(x,t)
\label{cc}
\eea
where we have used the reversed Heisenberg picture to denote our evolved position operator $\hat x(t)$. As before, we use a trigonometric expansion of the step function in the form
\bea
\theta(1/2-|\hat x(t)|) = \sum_{n,\pm}' \frac{-i}{k_n} e^{\pm ik_n(\hat x(t)\pm 1/2)} \equiv \sum_{n,\pm}' \frac{-i}{k_n} U_{n}^{\pm}
\label{expansion}
\eea
with a convenient abbreviation of the sum symbol given by

\bea
\sum_{n,\pm}' \equiv \sum_{n=0}^{\infty} \sum_{\pm} + 1/2 .
\eea

Finally, the wave can be written in terms of itself by noting that each exponential in the expansion is itself a unitary operator to be applied to the wave at the right. Such exponentials depend on the evolved position operator containing both $x$ and $p$ at $t=0$. In this way, many coordinate transformations can be induced on $\psi$:
\bea
\psi(x,t) = \sum_{n,\pm}' \frac{-i}{k_n} U_{n}^{\pm} \psi(x,t) = \sum_{n,\pm}' \frac{-i}{k_n} e^{i \xi_n(x,t)}\psi(\eta_n(x,t),t)
\eea
There are many choices of $\hat x(t)$ for which $U_{n}^{\pm}$ induces coordinate transformations. As the most general way (in two-dimensional phase space) take $\hat x(t)$ as a function of $x,p$ in the form

\bea
\hat x(t) = f_t(x) +\{g_t(x),p \}.
\label{evolve}
\eea
This operator induces a change of variables in $x$ and an affine transformation for each $n$ in the trigonometric sums above. To fulfill the initial data problem, we set $f_0(x)-ig_0'(x)=x$ and $g_0(x)=1/2$. Obviously, the function $\eta_n(x,t)$ producing the change of variables is to be determined for our evolution of choice (\ref{evolve}). Rotations, shearings and dilations can be obtained in this form.

What we have achieved so far can be summarized as follows: 

\begin{itemize}

\item The elements $G_t$ of an evolution group act on Hilbert space as $G_t : \hcal \rightarrow \hcal$. 

\item For $\psi \in \hcal$ with discontinuities in the form of pulses, one has $\psi = \Theta \psi$. The operator $\Theta$ is a projector. 

\item In the reversed Heisenberg picture one has $\Theta^{\small{(H)}} \equiv G_t \Theta G_{-t} $, which admits an expansion in terms of unitary operators $\Theta^{\small{(H)}}= \sum_n C_n U_n$.

\item For each $U_n$, a new transformation is produced: $x^{\small{(H)}} \equiv G_t x G_{-t} = f_t(x) + \{g_t(x),p\} $. The transformation is canonical, \ie $p^{\small{(H)}} = G_t p G_{-t}$. 

\item The overall result becomes $\psi(x,t)= \sum_n C_n \psi(\eta_n(x,t),t)$.

\end{itemize}

In the following we present some examples.

\section{Examples}

\subsection{Diffraction in a parametric harmonic oscillator}

Take as elements of the evolution group, the linear symplectic matrices with blocks $A,B,C,D$ acting on phase space $x,p$ \cite{quesne}. In the simplest two-dimensional application, these blocks can be considered scalars with the property $A(t)D(t)-B(t)C(t) =1$ and such that

\bea
\hat x(t) = A(t) x + B(t) p \\
\hat p(t) = C(t) x + D(t) p .
\eea
After substitution and the use of Baker-Campbell-Hausdorff formula, the replication formula becomes

\bea
\psi(x,t) &=& \sum_{n,\pm}' \frac{-i}{k_n} \exp\left(-ik_n^2 A(t)B(t)/2 \pm ik_n(A(t)x \pm 1/2)\right) \times \nonumber \\  &\times& \psi(x \pm k_n B(t),t)
\eea

\begin{figure}
\begin{center} 
 \begin{tabular}{ccc}
 \includegraphics[height=5.0cm, width=5.0cm]{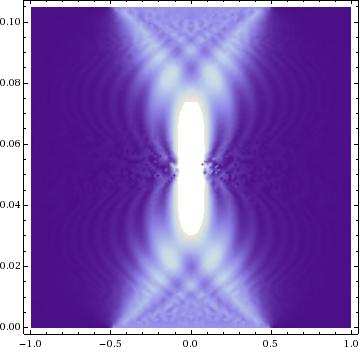} &
 \includegraphics[height=5.0cm, width=5.0cm]{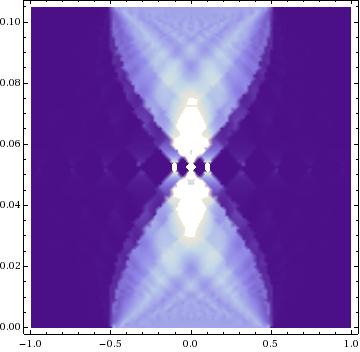} &
\includegraphics[height=5.0cm, width=5.0cm]{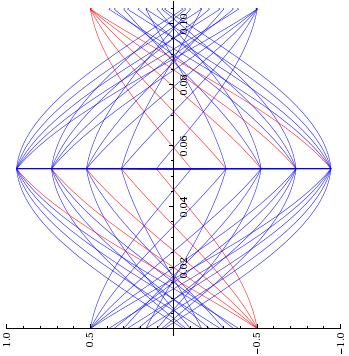}
 \end{tabular}
 \end{center} 
\caption{ \label{fig:osc} A comparison of diffraction patterns under a harmonic oscillator potential: Numerics, theory (with improved individual packets) and trajectories.}
 
 \end{figure}

For short times, we have an expansion in terms of building blocks or a {\it mother wave\ }, given by the initial condition:

\bea
\psi(x,t) &=& \sum_{n,\pm}' \frac{-i}{k_n} \exp\left(-ik_n^2 A(t)B(t)/2 \pm ik_n(A(t)x \pm 1/2)\right)\times \nonumber \\ &\times& \psi_0(x \pm k_n B(t)) + o(t^2)
\label{osc}
\eea
Furthermore, the short-time approximation can be improved by using a relation between harmonic and free evolution wavefunctions coming from the fact that both propagators are Gaussians. In the case of an initial condition given by the square packet, one obtains an interesting intensity pattern in agreement with the numerical evaluation. The trajectories are obtained from the phase factors in (\ref{osc}), parameterized by $A(t),B(t)$. For a time-independent harmonic oscillator, such functions are taken as $A(t)= \cos (\omega t), B(t) = - \sin(\omega t)/\omega $. See figure \ref{fig:osc} for a comparison.

\subsection{The emergence of wavelets}

Another example of interest emerges from the application of squeezing operations on wave packets. This shall lead to expansions similar to those of wavelet analysis \cite{daubechies}. Consider the transformation

\bea
\hat x(t) =  x + \frac{t}{2} \{x,p\} \\
\hat p(t) = \frac{1}{t} \log \left( tp + 1\right)
\eea
which is an unusual (but valid) example of a canonical, non-linear transformation generated by a hamiltonian which can be written down explicitly:

\bea
H= \frac{1}{2} \{ x  , (1-e^{-tp})/t^2 - p/t  \}.
\eea
The replication formula now reads

\bea
\psi(x,t) = \sum_{n,\pm}' \frac{(-)^n}{k_n} \exp\left(-i\pi (e^{\pm k_n t}-1)x/t \right) e^{\pm k_n t/2} \psi \left( e^{\pm k_n t /2} x,t \right) \nonumber \\ .
\eea
Which implies a true scaling of the wave packets by powers of the parameters. These are visible in the exponentials accompanying the argument $x$. Each power can be understood as a new scale factor 'zooming' into the pattern (this can be regarded as a microscope of increasing power). The function can be written for short times as 

\bea
\psi(x,t) = \sum_{n,\pm}' \frac{(-)^n}{k_n} \exp\left(-i\pi (e^{\pm k_n t}-1)x/t \right) e^{\pm k_n t/2} \psi_0 \left( e^{\pm k_n t /2} x \right) \nonumber \\ .
\eea
This is nothing else than a wavelet expansion, where the {\it mother wavelet\ } is given by the initial condition $\psi_0(x)$.

\subsection{Diffraction and the Gross-Pitaevskii equation}

\begin{center}
 \begin{figure}
 
 \begin{tabular}{lll}
 \includegraphics[height=5.0cm, width=5.0cm]{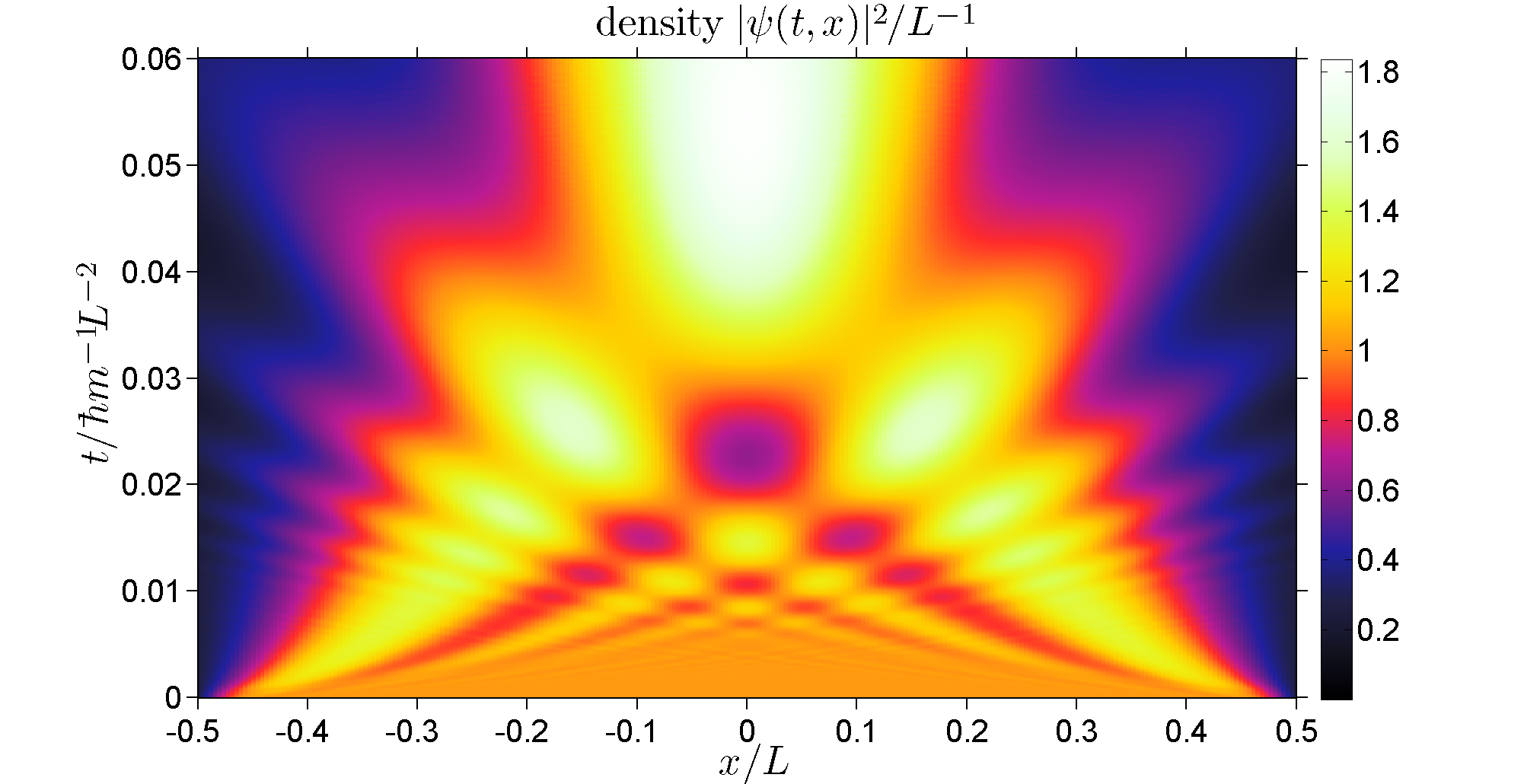} &
 \includegraphics[height=5.0cm, width=5.0cm]{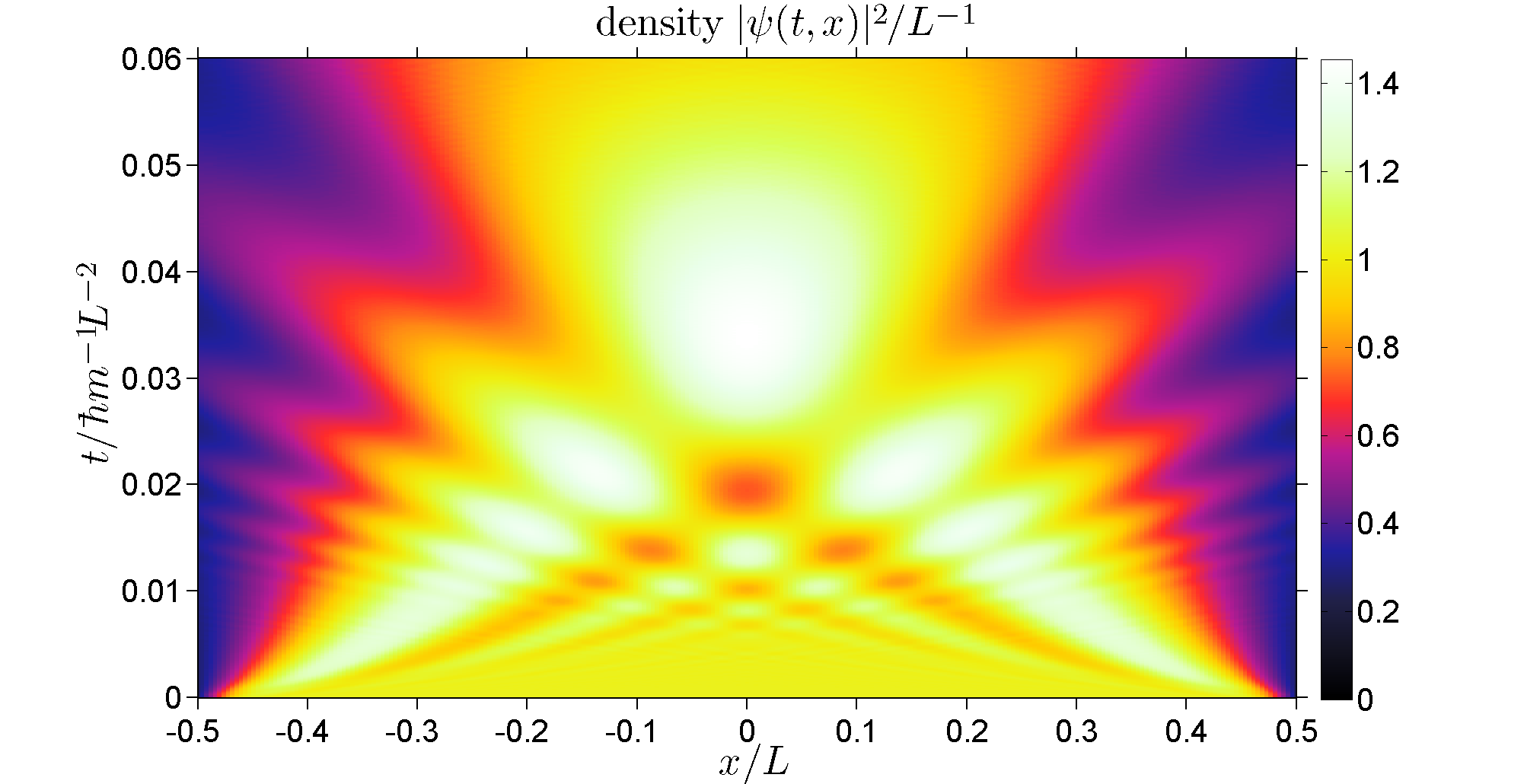} &
\includegraphics[height=5.0cm, width=5.0cm]{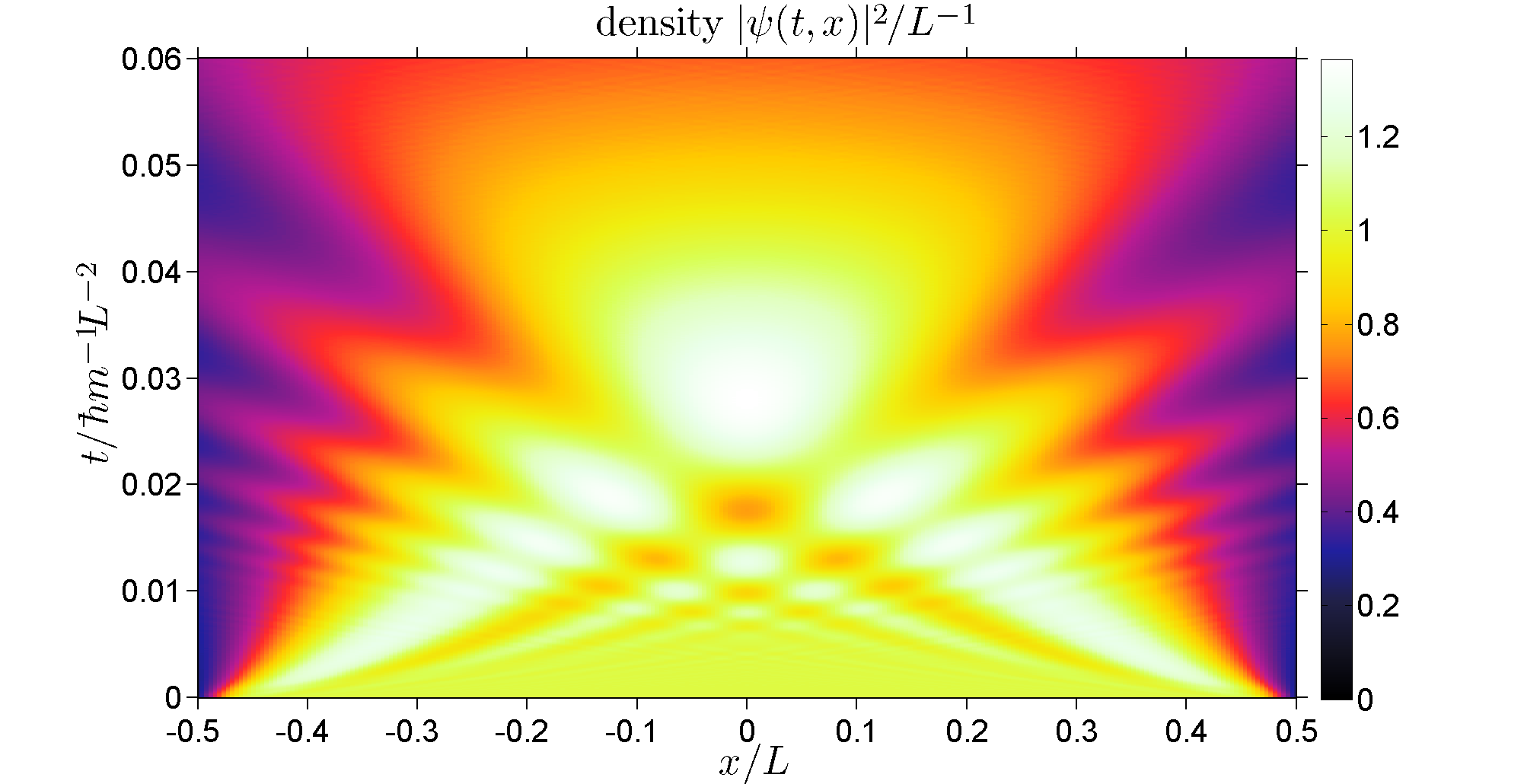}
 \end{tabular}
 
\caption{ \label{fig:gp} Probability density for increasing values of $g=0,50,100$. Courtesy of S. Arnold \cite{gp}.}
 
 \end{figure}
 \end{center}

Here, as a final example, we present the {\it numerical\ } solution to the evolution problem of a square packet governed by the Gross-Pitaevskii equation. 

\bea
\left[-\frac{1}{2}\nabla^2 + g|\psi(x,t)|^2 \right] \psi(x,t)= i \frac{\partial \psi(x,t)}{\partial t}, \qquad \psi(x,0) = \psi_0(x).
\label{gp}
\eea
We have chosen distributions with slightly smooth edges, showing thus that the  effect in question is robust. It is interesting to note that many of the features present in the linear Schr\"odinger equation, appear also in the non-linear case \cite{gp}. This occurs despite of the fact that in the numerical calculations, strong values of the coupling parameter $g=50,100$ have been used. See figure \ref{fig:gp}.

A brief explanation of the effect in question can be given in terms of the Gross-Pitaevskii equation plus the previously obtained replication formula for free evolution.
For $g=0$ we have
\bea
\psi(x,t) = \sum_{n=0}^{\infty} C_n e^{ik_n^2 t/2}e^{\pm ik_n(x\pm 1/2)} \psi_0(x \pm k_n t) + o(t^2).
\label{replication}
\eea
The non-linear potential can be incorporated in the evolution by means of the interaction picture. For short times and non-vanishing coupling,  we have that the corresponding operator can be replaced by a simple phase factor in the propagation problem. This is indicated in the following expression.
\bea
\psi_g(x,t) &=&  \int_{-\infty}^{\infty} dx' K(x,x';t,0) e^{-igt |\psi_0(x)|^2} \psi_0(x) + o(t^2) \nonumber \\ 
\label{grosspitaevskii}
\eea
which is in the form of Schr\"odinger propagation with an {\it effective\ } initial condition $e^{-igt |\psi_0(x)|^2} \psi_0(x)$. The intitially space-dependent phases produce additional focusing or defocusing in the previously obtained patterns (linear case). In the case of positive coupling $g$ we expect defocusing, preceded by a self-similar regime whose duration is shorter than in the free case.

\section{Concluding remarks}

In this contribution, devoted to the role of symmetries and self-similarity in diffraction patterns, we have shown that the Schr\"odinger evolution of discontinuities can be described by a wave function in terms of itself. This gave rise to a replicating pattern near the edges of an initial distribution. The notion of replication can be extended to other parabolic equations, including non-linear terms. A formal theory of the present treatment is to be developed: Envelope curves live in manifolds, rays in the exponents live in tangent spaces. It has been stressed that the symmetry groups producing the evolution in phase space {\it induce\ } transformations of wavepackets through the exponentiation of the evolved coordinate operators. 

It is desirable to find all these patterns in experimental setups, either using classical light, or the more ambitious manipulation of Bose-Einstein condensates. The current technology suggests \cite{jaouadi} that the needed edges in the initial distributions can be produced under special configurations of the confining electromagnetic traps
                    
\ack

The author is indebted to Stefan Arnold, William Case and Manuel Goncalves for fruitful discussions. Financial support from the DLR project QUANTUS is acknowledged.

\end{document}